# Ionic current rectification under concentration gradients and its application in evaluating surface charge properties of micropores

Long Ma,[1,2] Hongwen Zhang,[1,2] Bowen Ai,[1,2] Jiakun Zhuang,[1,2] Guanghua Du,[3] and Yinghua Qiu[1,2]*

1. Laboratory of High Efficiency and Clean Mechanical Manufacture of Ministry of Education, National Demonstration Center for Experimental Mechanical Engineering Education, School of Mechanical Engineering, Shandong University, Jinan, 250061, China

2. Shenzhen Research Institute of Shandong University, Shenzhen, 518000, China Key

3. Institute of Modern Physics, Chinese Academy of Sciences, Lanzhou, 730000, China

*Corresponding author: yinghua.qiu@sdu.edu.cn

## Abstract

Ionic current rectification (ICR) induced by electroosmotic flow (EOF) under concentration gradients can find many applications in micro/nanofluidic sensing and ionic circuits. Here, we focused on the cases with micropores of moderate length-diameter ratios, through experimental research and systematical simulations, the EOF-induced ICR was found to exhibit voltage-dependent ratios. In the considered cases with a weak EOF or strong ionic diffusion, a large deviation appears between the ion concentration inside the micropore and the bulk value, which fails the prediction by solution conductivity gradients. Based on our simulation results, effective equations were developed for the theoretical description of ion concentration distributions along the micropore axis under coupled concentration gradient and electric field. With the predicted ion distributions inside micropores, the ICR ratio can be conveniently calculated with the derived electrical resistance of the microfluidic system, which applies to micropores of 200 to 1000 nm in diameter. Because the surface charge density is the only unknown input parameter, our developed equations can be used to evaluate the surface charge density of micropores with the measured EOF-induced ICR ratio under concentration gradients.

## Introduction

Micro/nanopores offer a versatile platform for the investigation of mass transport,[1, 2] which have various applications in micro/nanofluidic sensing,[3-5] desalination,[6, 7] ionic circuits,[8-10] and energy conversion.[11, 12] In such confined spaces, surface charges on the pore walls significantly modulate the ion concentration distribution inside pores, as well as the ion and fluid transport through micro/nanopores.[1] Near charged pore walls, electric double layers (EDLs) form due to the electrostatic interaction between surface charges and free ions in solutions.[1, 13] With the application of electric fields across the micro/nanopore, counterions in EDLs have directional movement which can induce various useful phenomena, such as ionic selectivity of the pore,[14, 15] electroosmotic flow,[1, 16] ion concentration polarization,[17, 18] negative differential resistance,[19, 20] ionic current rectification,[9, 21-23] and memristive hysteresis.[23, 24]

Electroosmotic flow (EOF) is the fluid movement which is caused by the directional migration of ions inside EDLs based on the hydration effect of ions.[25, 26] Inside micro/nanopores, EOF can be strong enough to drag neutral nanoparticles or biomolecules through porous membranes.[27] In earlier reports, under a salt or viscosity gradient across micropores, EOF fills pores with the solution from the entrance side.[28, 29] Under opposite voltage biases, due to the conductivity difference in the solutions inside micropores filled by EOF, the ionic current through pores exhibits obvious rectification[27, 30-35] which is usually characterized by the ionic current rectification (ICR) ratio. With the mean-field theoretical predictions of the inner-pore resistance and access resistance, the EOF-induced ICR under viscosity or concentration gradients can be well understood.[21] Under a large conductivity gradient across the membrane, such as pure water at one side and 60% glycerol aqueous solution at the other side, for pores with a length-diameter ratio larger than ~100, the ICR ratio equals the conductivity ratio of solutions on both sides of the pore due to the ignored access resistance. In this case, ionic current rectifiers with a controlled ratio can be achieved.[27, 31] As the length-diameter ratio decreases, the ICR phenomenon becomes less obvious under viscosity or concentration gradients. For pores with a length-diameter ratio less than ~10, due to the ignored inner-pore resistance in the system, the ionic current rectification disappears which presents a rectification ratio equal to ~1.[21] In the cases with moderate length-diameter ratios under concentration gradients, the ICR ratio falls between 1 and the conductivity ratio of the solutions on both sides of the pore.[21]

EOF-induced ICR under concentration gradients can find many applications in micro/nanofluidic sensing and ionic circuits.[28, 31, 32, 35] Here, we focus on the development of theoretical prediction of the EOF-induced ICR ratio through micropores with moderate length-diameter ratios under concentration gradients. Through both microfluidic experiments and COMSOL multiphysics simulations, the current-voltage curves are obtained which show obvious ionic current rectification but voltage-dependent ICR ratios. With the consideration of factors modulating the electroosmotic flow inside micropores, the distributions of ionic concentration along the pore axis have been investigated and predicted theoretically. Then the EOF-induced ICR ratios can be fitted well with the derived equations of the ion concentration distribution along the pore axis. Further, a simple method is developed for the evaluation of the surface charge density on pore walls, which serves as an important parameter of porous membranes. Our method only requires one input parameter, i.e. the EOF-induced ICR ratio obtained under a concentration gradient. Based on the similar concentration distribution inside micropores with a diameter varying from 200 nm to 1000 nm, our developed equation can be used for the surface charge density prediction under those pore sizes.

## Experiment and Simulation Details

In the microfluidic experiments, polyethylene terephthalate (PET) micropores were used which were prepared by the track-etching technique.[24, 36, 37] Heavy ion bombardment was employed to introduce latent tracks into the PET membranes with a thickness of 12 μm (GSI, Darmstadt, Germany, and Lanzhou HIRFL, China).[36, 38] Before wet etching, both sides of the PET membranes underwent irradiation with the ultraviolet light for one hour (UVP UVGL−25, CA, USA). Later, the membrane was etched in 0.5 M NaOH solution at 70 °C (Julabo VIVO B3, Beijing, China). Regulation of the etching time allowed for the generation of pores with controlled diameters. After the fabrication of the PET pore, the diameter was characterized by the ionic conductance at 1 M KCl (Figure S1). In the later experiments, salt gradients were applied across the micropores with KCl solutions. 50 mM KCl was set at the low-concentration side. The concentration on the high-concentration side was considered from 50 mM to 1 M. All solutions were buffered to pH 10 using 10 mM Trizma base.[28, 39] All chemicals used in the experiment were purchased from Sigma−Aldrich. Deionized water was produced by Direct−Q 3UV (18.2 MΩ, MilliporeSigma, Burlington, USA). A conductivity meter (Mettler Toledo FE 38, Shanghai, China) facilitated the measurement of solution conductivities. Picoammeters (Keithley 6487, Keithley Instruments, Solon, USA, and MeZero, Meili Nanopore Tech, Shenzhen, China) were used to record the ionic currents through micropores. As shown in Figure 1a, the working and ground electrodes were placed in the high- and low-concentration reservoirs, respectively.

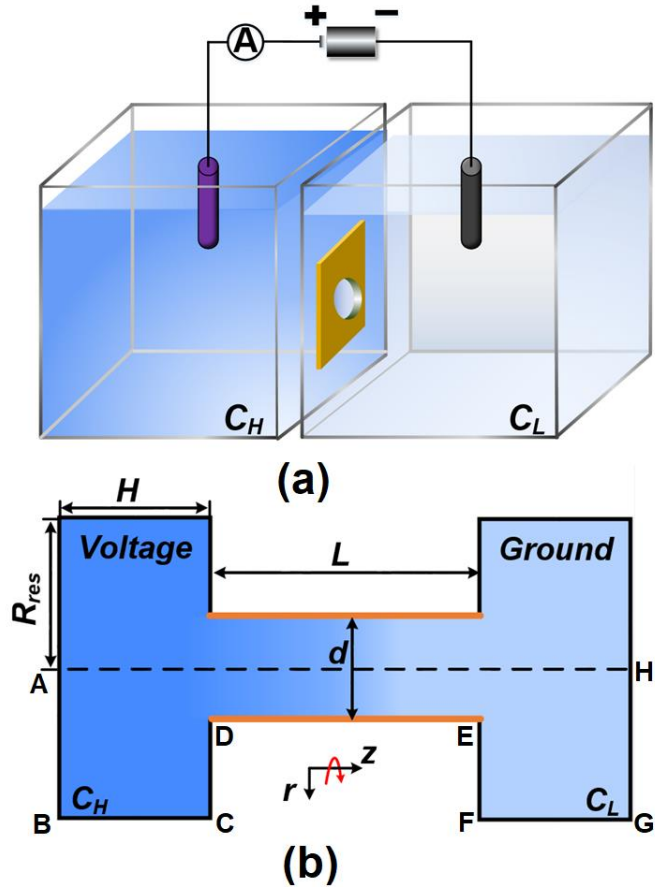

Figure 1 Schematic diagrams of the experimental setup (a) and the simulation model (b, not to scale). $C_H$ and $C_L$ denote the high and low concentrations of the solutions on both sides of the pore. The radius and length of the reservoir, as well as the diameter and length of the pore, are denoted as $R_{res}$, $H$, $d$, and $L$, respectively. Charged pore walls are shown by orange lines. Letters A-H are the edge points to show the simulation geometry.

Ion transport and fluid flow inside micropores were simulated with COMSOL Multiphysics through coupled Poisson–Nernst–Planck (PNP) and Navier–Stokes (NS) equations (Eqs. 1-4).[40][41] The PNP equation describes the distributions of the electrical potential and ion distribution, as well as the ion transport in the system. The NS equation describes the fluid flow in the nanopore and reservoirs.

$$\varepsilon \nabla^2 \varphi = -F \sum_{i=1}^{N} z_i C_i \quad (1)$$

$$\nabla \boldsymbol{J}_i = \nabla \left( -D_i \nabla C_i + \boldsymbol{u} C_i - \frac{F z_i C_i D_i}{RT} \nabla \varphi \right) = 0 \quad (2)$$

$$\mu \nabla^2 \boldsymbol{u} - \nabla p - \sum_{i=1}^{N} (z_i F C_i) \nabla \varphi = 0 \quad (3)$$

$$\nabla \boldsymbol{u} = 0 \quad (4)$$

where $\varepsilon$, $\nabla$, $\varphi$, $F$, and $N$ are the dielectric constant of aqueous solutions, gradient operator, electrical potential, Faraday constant, and number of ionic species in solutions. $C_i$, $z_i$, $\boldsymbol{J}_i$, and $D_i$ are the concentration, valence, ionic flux, and diffusion coefficient of ionic species $i$ (including cations and anions) in solutions. $R$, $T$, and $\boldsymbol{u}$ are the temperature, gas constant, and velocity of the fluid, respectively. $p$ and $\mu$ are the pressure and viscosity of aqueous solution.

As shown in Figure 1b, a micropore connects the high- and low-concentration reservoirs, whose radius and height were set to 5 μm. The pore length ($L$) was changed from 2 μm to 12 μm, with a default length of 12 μm. The diameter ($d$) of the pore was varied from 200 nm to 1000 nm, with a default value of 475 nm corresponding to the diameter of a PET micropore used in the microfluidic experiments. During the simulation, diffusion coefficients of K$^+$ and Cl$^-$ ions in KCl solutions were set to 1.96×10$^{-9}$ and 2.03×10$^{-9}$ m$^2$/s, respectively.[42] The solution concentration in the low-concentration reservoir was consistently maintained at 50 mM. By adjusting the concentration of the high-concentration reservoir, the salt gradient across the micropore was tuned from 2 to 20, with a default value of 20. The activity coefficients of KCl solutions with different concentrations were considered, which are shown in Table S1.[43] The surface charge density varied from −0.02 C/m$^2$ to −0.04 C/m$^2$ with a default value of −0.025 C/m$^2$. The default surface charge density was selected to obtain the best fit of simulation results to the experimental results with the PET micropore. The applied voltage was changed from −2 V to 2 V. The temperature and dielectric constant of the aqueous solution were set to 298 K and 80, respectively. The boundary conditions are listed in Table 1. The meshing strategy was consistent with our previous studies,[23, 40, 41, 44] which is shown in Figure S2.

**Table 1**. Boundary conditions used in numerical modeling. Coupled Poisson-Nernst-Planck and Navier-Stokes equations were solved with COMSOL Multiphysics.

| Surface | Poisson | Nernst-Planck | Navier-Stokes |
|---|---|---|---|
| AB | constant potential (Ground) $\phi=0$ | constant concentration $c_i=C_H$ | constant pressure $p=0$ no viscous stress $\mathbf{n}\cdot[\mu(\nabla\mathbf{u}+(\nabla\mathbf{u})^T)]=0$ |
| BC, FG | no charge $-\mathbf{n}\cdot(\varepsilon\nabla\phi)=0$ | no flux $\mathbf{n}\cdot\mathbf{N}_i=0$ | no slip |
| HG | constant potential $\phi=V_{app}$ | constant concentration $c_i=C_L$ | constant pressure $p=0$ no viscous stress $\mathbf{n}\cdot[\mu(\nabla\mathbf{u}+(\nabla\mathbf{u})^T)]=0$ |
| AH | axial symmetry | axial symmetry | axial symmetry |
| CD, DE, EF | $-\mathbf{n}\cdot(\varepsilon\nabla\phi)=\sigma_w$ | no flux $\mathbf{n}\cdot\mathbf{N}_i=0$ | no slip $\mathbf{u}=\mathbf{0}$ |

$\phi$, $\varepsilon$, $C_L$, $C_H$, $p$, $\mathbf{n}$, $\mathbf{N}_i$, $\mathbf{u}$, $\sigma_w$, $\mu$ are the surface potential, dielectric constant, low bulk concentration, high bulk concentration, pressure, normal vector, flux of ions, fluid velocity, surface charge density of the pore wall, solution viscosity, respectively.

The ionic flux is the flux of cations or anions through the cross-section of the nanopore, which can be obtained conveniently at the simulation boundaries. Then, the ionic current ($I$) at different voltages was calculated by integrating the cation and anion fluxes at the reservoir boundary with Eq. 5.[40, 45]

$$I = \int_S F\left(\sum_i^2 z_i \mathbf{J}_i\right) \cdot \mathbf{n} \, dS \tag{5}$$

where $S$ represents the reservoir boundary, and $\mathbf{n}$ is the unit normal vector.

## Results and Discussion

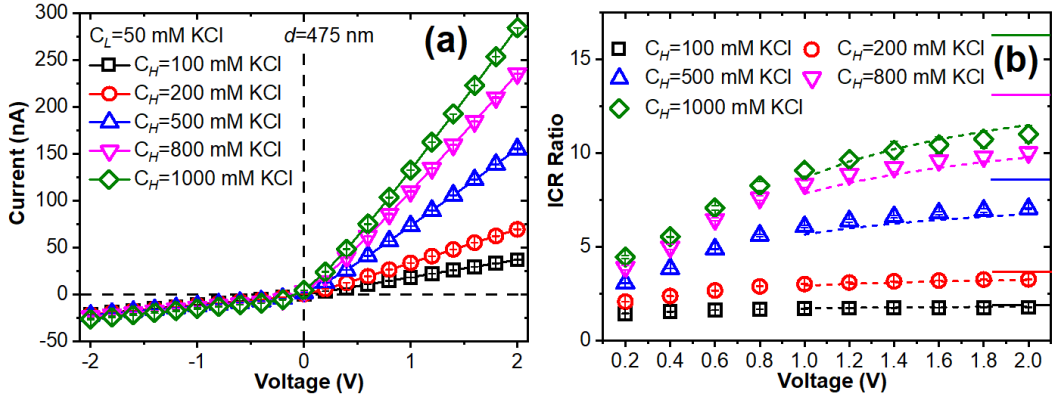

Figure 2 Experimental ionic current behaviors obtained with a PET micropore with 475 nm in diameter and 12 μm in length. (a) Current-voltage (I-V) curves and (b) ICR ratios across the micropore at different voltages calculated with $|I_+|/|I_-|$. Dashed lines represent the resistance ratio with Eqs. 8 to 11 based on simulated ion concentrations along the pore axis. Solid lines represent the conductivity ratio of high- and low-concentration solutions.

Figure 2a shows the I-V curves obtained with a PET micropore of 475 nm in diameter and 12 μm in length under different concentration gradients. In pH 10 solutions, the PET membrane is negatively charged. With the setup of electrodes shown in Figure 1a, at negative voltages, counterions i.e. K$^+$ ions migrate from the low-concentration side to the high-concentration side. Due to the hydration effect of ions, the directional movement of counterions inside electric double layers can induce the directional fluid flow, i.e. the formation of EOF.[25, 26] The induced EOF fills the micropore with the low-concentration solution, i.e. 50 mM KCl, which results in an almost unchanged current at various gradients.[28] While at positive voltages, EOF in the reversed direction drags the high-concentration solution into the micropore which causes a much larger ion current. Our results exhibit the EOF-induced ionic current rectification, which has been reported in several previous works.[21, 28, 31] As shown in Figure 2b, the ICR ratios present a voltage-dependent trend. With the increase of the applied voltage across the pore, the ICR ratio has a sharp increase first and then approaches its saturation linearly with a much smaller slope.

In aqueous solutions, the EDL thickness is inversely related to the salt concentration.[13] For the cases with smaller concentration gradients than 4, thicker EDLs can result in stronger EOF that can fill the pore with the entrance solution more effectively (see below). The solution conductance inside micropores under both positive and negative voltages is close to the bulk conductance of the high- and low-concentration solutions. Then, smaller concentration gradients facilitate a quicker stabilization of the ICR ratio. The ICR ratio at ±2 V can be approximated by the conductivity ratio of solutions across the membrane. However, under a higher concentration gradient than 2, deviation starts to appear between the ICR ratio at 2 V and the solution conductivity ratio, which can be 48% for the case with a concentration gradient of 20. The characteristics of ICR ratios shown in Figure 2b were also confirmed by two other PET micropores (Figure S3).

Then, we focus on the theoretical prediction of the ICR ratio in pores with moderate length-diameter ratios under concentration gradients. For the PET micropore used in this work, the length-diameter ratio is ~25, much smaller than 100.[21] Based on the mean-field theory, the access resistance ($R_{ac}$) and inner-pore resistance ($R_p$) can be described with Eqs. 6 and 7, respectively, which are mainly determined by micropore dimensions and the conductivity of the solution.[21]

$$R_{ac} = \frac{1}{2\kappa d} \quad (6)$$

$$R_p = \frac{4L}{\pi \kappa d^2} \quad (7)$$

$$\text{ICR ratio} = |I_+/I_-| = R_-/R_+ \quad (8)$$

$$R_\pm = 2R_{ac(\pm)} + R_{p(\pm)} = \frac{1}{\kappa_{(\pm)} d} + \frac{4L}{\pi \kappa_{(\pm)} d^2} \quad (9)$$

where $d$ and $L$ are the diameter and length of the pore, $\kappa$ is the conductivity of the solution, and the subscript symbols +, −, and ± denote the applied voltage polarity.

Under the assumption that EOF fills the pore with the entrance solution, the ICR ratio can be predicted with the resistance ratio of the whole system based on the solution distribution under opposite bias polarities (Eqs. 8 and 9).[21] However, for micropores with a length-diameter ratio falling between ~10 and ~100, the roughly predicted ICR ratio falls between 1 and the conductivity ratio of solutions on both pore sides, which has no dependence on the applied voltage.

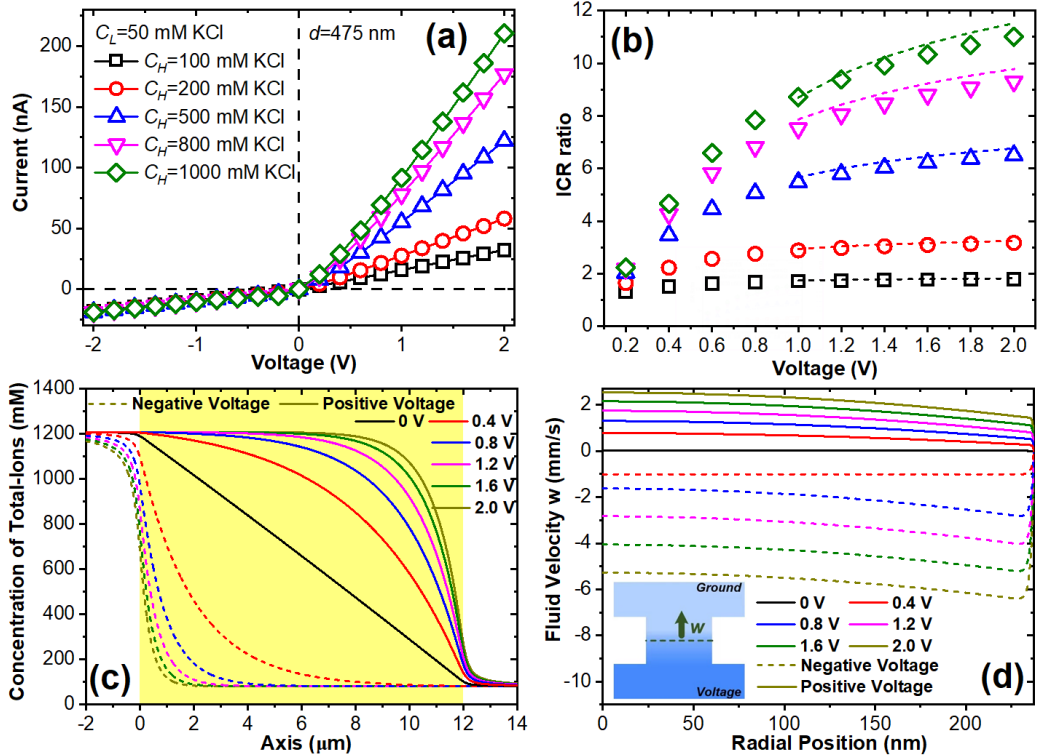

Figure 3 Simulation results of ionic current behaviors and total-ion concentration distributions with a micropore with 475 nm in diameter and 12 µm in length. (a) I-V curves. (b) ICR ratios at different voltages. Dashed lines represent the resistance ratio with Eqs. 8 to 11 based on simulated ion concentrations along the pore axis. (c) Concentration distributions of ions along the pore axis at different voltages. The salt gradient is 50:1000 mM KCl. (d) Radial distributions of EOF velocity at the central cross-section of the pore. The inset shows the location of the central cross-section. The surface charge density of pore walls was set at −0.025 C/m$^2$.

Microfluidic simulations were conducted to investigate the detailed physics behind the voltage-dependent ICR ratio. As shown in Figure 3, simulated I-V curves (Figure 3a) were obtained which present almost the same ICR ratios (Figure 3b) as those in the experiments. Figure S4 shows the comparison between ICR ratios from both experiments and simulations. Due to the positive dependence of solution conductivity on salt concentration,[46] the concentration distribution of total ions along the axis was explored at various voltages under 50:1000 mM KCl (Figure 3c).[44] At 0 V, the ion concentration diminishes linearly across the pore from the high concentration to the low concentration. In this case, ion transport is governed by the process of diffusion. With the application of a voltage across the pore, the pore is gradually filled with the bulk solution at the entrance side by the induced EOF. At positive voltages, both EOF and diffusive flow drive the high-concentration solution into the pore, which fills the pore with the entrance solution more efficiently at higher voltages. At the pore exit where in contact with the low-concentration solution, thicker EDLs enhance the transport of counterions along charged pore surfaces, which decreases the inner-pore concentration.[45] The relatively large electric field strength induced by the low concentration near the pore exit (Figure S5) further accelerates the migration of counterions out of the micropore. From Figure S6, the vortex formed at the micropore exit can draw the low-concentration solution into the pore, which also decreases the ion concentration inside the micropore.

After the application of negative voltages, the low-concentration solution is filled into the micropore by a relatively stronger EOF (Figure 3d) compared with that at positive voltages due to the thicker EDLs in diluter solutions. Because of the opposite directions of the concentration gradient and EOF, free ions in high-concentration solutions diffuse into the micropores at low voltages, which increases the inner-pore concentration at the pore exit. As the voltage increases, the enhanced ion migration and EOF (Figure 3d) gradually inhibit the ion diffusion under concentration gradients, which fill the micropore with the low-concentration solution to a greater extent. Also, due to the migration of $Cl^-$ ions and the ion diffusion, the micropore cannot be full of the low-concentration solution. Compared with the cases at positive voltages, the turning points of ion concentration profiles inside the micropore are located closer to the pore entrance under negative voltages.

Under the coupled electric field and concentration gradient, the ion transport through the pore has three components, i.e. ion diffusion, ion migration, and convection-induced ion transport. To display the three contributions to the total current, additional simulations were conducted without the consideration of the NS equation, i.e. simulations only considering ion diffusion and migration.[47] From Figure S7, without the consideration of fluid flow inside the system, the ICR phenomenon disappears. The diffusion contribution to the current under salt gradients is very little. For the current values obtained from simulations without the NS equation, the current is contributed almost by the ion migration. Then, the current contribution by the EOF can be obtained by subtracting the current values with the consideration of the NS equation from those without the consideration of the NS equation. As shown in Figure S8, we obtained the current contributions from ion diffusion, ion migration, and EOF-induced ion transport under different voltages. The diffusion current is very little. Note that due to the larger diffusion coefficient of $Cl^-$ ions than $K^+$ ions, the total diffusion current is negative. Under electric fields, the ion migration has a considerable contribution to the current. While the current at the same voltage but opposite polarities share similar values. For the EOF-induced ion transport, under both polarities, it contributes to the total current positively. Note that in the micropore EOF is induced by the migration of cations. Considering the respect migration direction of cations and anions, the EOF enhances and prohibits the migration of $K^+$ ions and $Cl^-$ ions under both positive and negative voltages.[47]

Equation 10 describes the solution conductivity which is directly related to the salt concentration, as well as the valence and diffusion coefficient of ions. In KCl solutions, both contributions from $K^+$ and $Cl^-$ ions to the conductivity are included.

$$\kappa_{(\pm)} = \frac{F^2}{RT} \sum |z_i| C_{i(\pm)} D_i \tag{10}$$

$$R_{p(\pm)} = \int \frac{4L}{\pi \kappa_{(\pm)} d^2} \tag{11}$$

where $i$ represents the ionic species $i$ ($K^+$ ions and $Cl^-$ ions). $C_i$, $D_i$, and $z_i$ are the concentration, diffusion coefficient, and valence of ionic species $i$. $C_{i(\pm)}$ is the ion concentration of $K^+$ or $Cl^-$ ions under a positive or negative voltage. $F$, $R$, and $T$ are the Faraday's constant, gas constant, and temperature.

In the system with a micropore, the pore resistance can be integrated by the binning method[48] over the whole pore length (Eq. 11). Based on the simulated concentration distributions along the pore axis shown in Figure 3c, the system resistance was calculated under both biases with coupled Eqs. 8, 10, and 11. The calculated ICR ratios under various voltages are shown in Figure 3b by the short-dashed line, which captures the trend and values of simulation ICR ratios at various voltages.

From the above results, ICR ratios across the micropore with a moderate length-diameter ratio of ~25 can be predicted with the actual concentration distribution along the pore (Figure 3b). Inside micropores, EOF is the main motivation to modulate the concentration distribution, whose velocity can be affected by the micropore dimensions, salt concentration, applied voltage, and surface charge density (Eq.12).[47]

$$v_{EOF} \approx 205.6 \frac{\varepsilon V}{\eta(4L+\pi d)} \mathrm{arcsinh}(\frac{\sigma}{0.117\sqrt{C}}) \qquad (12)$$

in which $v_{EOF}$, $V$, $C$, $\sigma$, $\eta$ are the velocity of EOF, applied voltage, the concentration of solutions, surface charge density, and viscosity of the fluid, respectively.

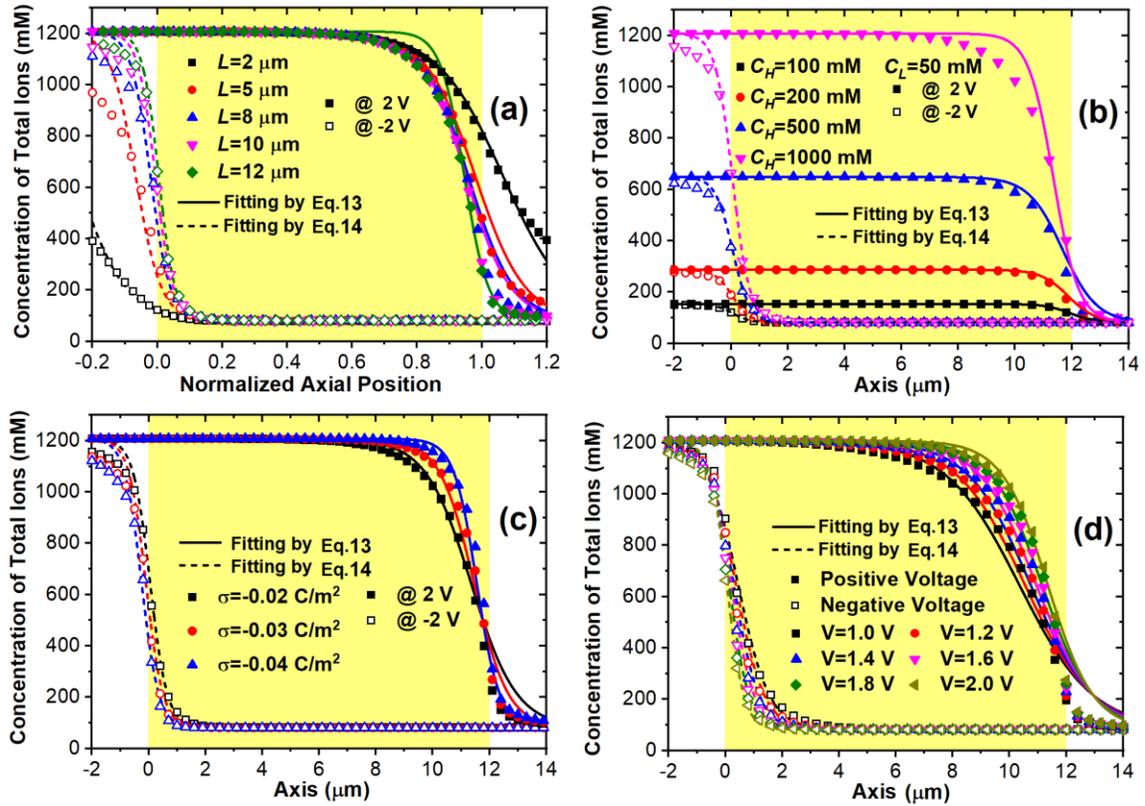

Figure 4 Prediction for the concentration distributions of ions along the pore axis under various conditions, such as different pore lengths (a), concentration gradients (b), surface charge densities (c), and voltages (d). Short-dashed lines and solid lines are the theoretical fittings with Eqs. 13 and 14. The default pore length, salt gradient, surface charge density, and voltage are 12 μm, 50:1000 KCl, −0.025 C/m$^2$, and ±2 V, respectively. The pore diameter was 475 nm.

Based on our systematic simulations, we attempted to derive the theoretical prediction of the voltage-dependent concentration distribution along the pore axis. Figure 4 shows the distribution of ion concentration (symbols) in the cases with different conditions such as different pore lengths, salt gradients, surface charges, and applied voltages. Note that we also considered the effect of the pore diameter on the ICR ratio.[1, 49] From Figure S9, for micropores with a diameter varying from 200 to 1000 nm, the ICR ratio under salt gradients remains almost constant which is due to the similar distribution of ion concentration. Thus, the diameter influence can be ignored.

In the microfluidic system, the pore length determines the electric field strength and the salt gradient, which affect the EOF velocity and ionic diffusion, respectively.[40, 44] Figure 4a exhibits the concentration distributions along the normalized pore length at ±2 V. Under a concentration gradient of 50:1000 mM KCl, ion concentrations share a similar trend inside micropores with a length varying from 2 to 12 μm. At similar turning points, the concentration changes from ~1200 mM to ~100 mM. Note that in the simulations we considered solution activities.[43] In all cases, the main difference is the slope of the concentration change at the pore exit, with shallower slopes appearing in short micropores. This may be attributed to the strong ion diffusion at small pore lengths.

With 50 mM KCl as the low-concentration solution, the concentration gradient across the micropore affects the strength of ionic diffusion and the Debye length inside the pore.[45] At a larger salt gradient, due to the lower surface potential at a higher ion concentration, the EOF speed inside micropores decreases at 2 V, which induces a farther turning point on the concentration profile away from the pore exit. While at a lower concentration gradient, stronger EOF can fill the pore with the high-concentration solution more effectively which leads to an ICR ratio closer to the conductivity ratio of bulk solutions. At −2 V, though concentration gradients have an opposite direction to the EOF, due to the weak ion diffusion through long micropores, concentration profiles share a similar turning point with the unchanged low-concentration solution.

Both surface charge density and applied voltage are important parameters, that positively correlate to the speed of EOF (Eq. 12).[47] As shown in Figures 4c and 4d, a higher surface charge density and voltage can induce stronger EOF which drives the turning points of the concentration profile closer to the pore exit.

Under both positive and negative voltages, solutions from the high- and low-concentration reservoirs fail to fill the entire micropore. Then, the description of the ion concentration distribution along the micropore is useful for the prediction of the ICR ratio under concentration gradients. Based on our results from simulations (Figure 4), Eqs.13 and 14 are proposed to quantitatively describe the total ion concentration along the pore axis under various conditions, including different pore lengths, concentration gradients, surface charge densities, and voltage biases. Considering that the concentration distribution is determined by the salt gradient and the induced EOF, all related parameters were included in our proposed equations.[47] As shown in Figure 4, all the concentration distributions under various conditions along the pore axis can be theoretically predicted by Eqs. 13 and 14. Please note that the predicted concentration by Eq. 13 or 14 at the center of the cross-section of the pore entrance or exit was applied for the value of $C_i$ in the calculation of the corresponding $R_{ac}$. Due to the application of micropores, both anions and cations share the same concentration distribution along the pore axis (Figure S10). The concentration of cations or anions is half of the values predicted by Eq. 13 or 14.

$$C_+ = \frac{2\alpha_H C_H - 2\alpha_L C_L}{1+e^{\frac{x-(0.922L-0.65C_H+10000|\sigma|+1000V-1264)}{0.06L+0.4C_H-20000|\sigma|-500V+1080}}} + 2\alpha_L C_L \qquad (13)$$

$$C_- = \frac{2\alpha_H C_H - 2\alpha_L C_L}{1+e^{\frac{x-(0.055L-15000|\sigma|+400V+490)}{0.015L+250V+620}}} + 2\alpha_L C_L \qquad (14)$$

where $\alpha_H$ and $\alpha_L$ are the activity coefficients of solutions on the high- and low-concentration sides. $x$ is the axial position inside the micropore.

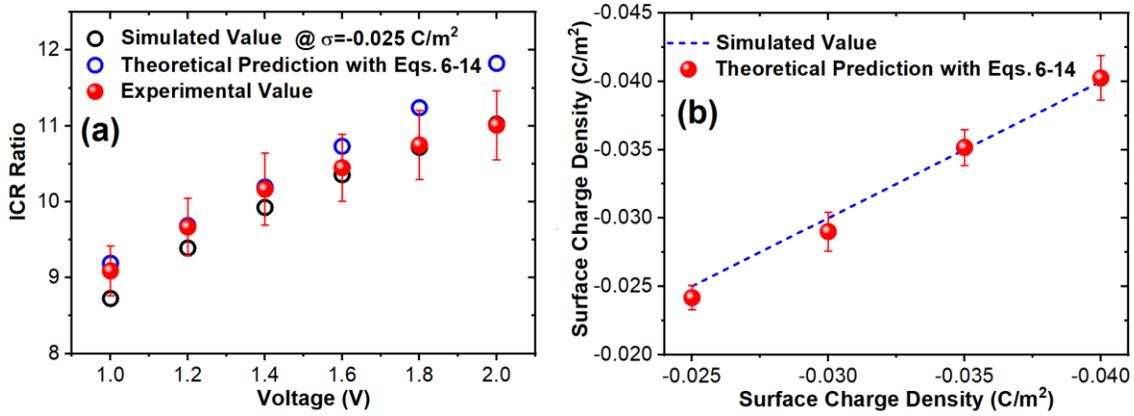

Figure 5 Theoretical prediction of the ICR ratio (a) and surface charge density (b) of micropores. Error bars represent the standard deviation of the surface charge density obtained at six different voltages from 1 to 2 V. The length and diameter of the micropore are 12 μm and 475 nm, respectively. The salt gradient is 50:1000 mM KCl.

Based on the developed theoretical description of ion concentrations inside micropores, system resistances can be evaluated with derived solution conductivity, and provide a quantitative prediction of ICR ratios.[44] Figure 5a shows the simulated and experimental values of the ICR ratio as well as the theoretical predictions under the concentration gradient of 20 folds. Note that due to the strong ionic diffusion under the concentration gradient, we mainly focus on the applied voltage greater than 1 V. With the surface charge density of −0.025 C/m$^2$, the simulated ICR ratios agree well with the experimental data. From Eqs. 13 and 14, with the surface charge density as one input parameter, the theoretical predictions of ICR ratios based on ion concentrations can be used to fit the simulated and experimental values (Figure 5a). After the determination of the surface charge density, Eqs. 13 and 14 can provide a good theoretical prediction for the experimental ICR ratios under various gradients. Please note that we also considered a case with a 100-nm-in-diameter nanopore (Figure S11). In this case, the deviation between simulation results and our developed theoretical prediction reaches above 10%.

Here, an evaluation method for the surface charge density can be raised based on the experimental ICR ratios and our developed equations of the concentration distribution. With the experimental ICR ratio, pore dimensions, applied voltage, and solution concentrations as input parameters, the surface charge density can be predicted from coupled Eqs. 6-14. As shown in Figure 5b, the predicted surface charge density can be obtained with the simulated ICR ratios under 50:1000 mM KCl solutions. The theoretical prediction of the surface charge density with our developed equations exhibits good agreement with the values used in simulations. We also considered the cases with NaCl solutions. The diffusion coefficients of Na$^+$ and Cl$^−$ ions in NaCl solutions were set to 1.33×10$^{−9}$ and 2.03×10$^{−9}$ m$^2$/s, respectively.[42] As shown in Figure S12, our developed equations are also applicable to NaCl solutions in which cations and anions have different diffusion coefficients. Our method may provide an easy prediction of the surface charge density with only one experimental ICR ratio under a salt gradient. Due to the unchanged ICR ratios and concentration distributions inside micropores with diameters from 200 to 1000 nm, our methods can be used for the evaluation of the surface charge density for micropores at this diameter range.

## Conclusions

Under electric fields, EOF can fill micropores with solutions on the pore entrance, which induces the ionic current rectification under concentration gradients. For micropores with moderate length-diameter ratios, EOF-induced ICR ratios exceed 1 but are less than the conductivity ratio of the high- and low-concentration solutions, because the entrance solution driven by EOF cannot fill the entire micropore. Here, systematical simulations have been conducted to investigate the ion distribution inside micropores. The system resistance can be evaluated which provides a viable alternative for the prediction of the ICR ratio. From simulations under various conditions, in the cases with a small voltage, a low surface charge density, or a high concentration gradient, due to the lower EOF velocity or stronger ionic diffusion, the ion concentration inside the micropore has a larger deviation from the bulk value. Then, a theoretical description of the ion concentration distributions inside micropores is developed, which can be used to predict the ICR ratios under various conditions for micropores at the diameter range from 200 to 1000 nm. With the surface charge density as the only unknown parameter, a simple evaluation method of the surface charge density is proposed with our developed equations and one experimental ICR ratio under a salt gradient.

## Supporting information

See supplementary material for simulation details and additional simulation results.

## Acknowledgment

This research was supported by the National Natural Science Foundation of China (52105579), the Guangdong Basic and Applied Basic Research Foundation

## AUTHOR DECLARATIONS

Conflict of Interest

The authors have no conflicts to disclose.

Author Contributions

**Long Ma:** Simulation, Investigation, Data curation, Writing - original draft, review & editing. **Hongwen Zhang:** Data curation. **Bowen Ai:** Data curation. **Jiakun Zhuang:** Data curation. **Guanghua Du:** Investigation. **Yinghua Qiu:** Resources, Conceptualization, Methodology, Writing - original draft, review & editing, Supervision, Funding acquisition.

## DATA AVAILABILITY

The data that support the findings of this study are available from the corresponding author upon reasonable request.